\documentclass[10pt, a4paper, twocolumn, aps, pra, showpacs, longbibliography, superscriptaddress]{revtex4-1}
\pdfoutput=1
\usepackage[utf8x]{inputenc}
\usepackage{ucs}
\usepackage{amsmath}
\usepackage{amsfonts}
\usepackage{amssymb}
\usepackage{makeidx}
\usepackage{cellspace,booktabs}
\usepackage{natbib}
\usepackage{lipsum}
\usepackage{bm}
\usepackage{bbm}

\usepackage{color}
\definecolor{light-gray}{gray}{0.55}

\usepackage{microtype}

\usepackage{graphicx}

\renewcommand{\dag}{^{\dagger}}
\newcommand{\exv}[1]{ \langle #1 \rangle }

\newcommand{\bra}[1]{ \langle #1 \rvert }
\newcommand{\ket}[1]{ \lvert #1 \rangle}
\newcommand{\braket}[2]{\langle #1 \vert #2 \rangle }

\newcommand{\tr}[2][]{\text{Tr}_{ #1 } ( #2 )}

\begin{document}

\begin{abstract}

We propose a scheme for quantum teleportation between two qubits, coupled sequentially to a cavity field.  An implementation of the scheme is analyzed with superconducting qubits and a transmission line resonator, where measurements are restricted to continuous probing of the field leaking from the resonator rather than instantaneous projective Bell state measurement. We show that the past quantum state formalism [S. Gammelmark et al, Phys. Rev. \textbf{111}, 160401] can be successfully applied to estimate what would have been the most likely Bell measurement outcome conditioned on our continuous signal record. This information determines which local operation on the target qubit yields the optimal teleportation fidelity. Our results emphasize the significance of applying a detailed analysis of quantum measurements in feed-forward protocols in non-ideal leaky quantum systems.

\end{abstract}

\date{\today}
\author{Eliska Greplova}
\thanks{E-mail: eliska.greplova@phys.au.dk}
\affiliation{Department of Physics and Astronomy, Aarhus University, DK-8000 Aarhus C, Denmark}
\author{Klaus Mølmer}
\thanks{E-mail: moelmer@phys.au.dk}
\affiliation{Department of Physics and Astronomy, Aarhus University, DK-8000 Aarhus C, Denmark}
\author{Christian Kraglund Andersen}
\thanks{E-mail: ctc@phys.au.dk}
\affiliation{Department of Physics and Astronomy, Aarhus University, DK-8000 Aarhus C, Denmark}

\title{Quantum teleportation with continuous measurements}

\maketitle

\section{Introduction}
Quantum teleportation \cite{PhysRevLett.70.1895} is a protocol allowing the application of non-local quantum superposition states in quantum information \cite{nielsen2010quantum}, computation \cite{gottesman1999demonstrating}, and cryptography \cite{RevModPhys.74.145}. Quantum teleportation has been experimentally demonstrated in a number of systems \cite{bouwmeester1997experimental, furusawa1998unconditional, riebe2004deterministic, barrett2004deterministic, sherson2007, steffen2013deterministic, pfaff2014unconditional}, and it allows two parties, $A$ and $B$, who share a maximally entangled state, to 'teleport' an unknown quantum state, $\ket{\psi}$, from the location of $A$ to the location of $B$ using only local operations and classical communication. Quantum teleportation is also refereed to as disembodied transfer, emphasizing that no properties of the teleported quantum state are at any time detectable in the spatial region between the locations A and B.

We propose a scheme for teleportation between qubits, where we use a cavity field as the communication channel, see Fig.~1. After preparation of an entangled state of the cavity field and qubit $B$, the qubit is detuned away from the cavity resonance and no longer interacts with the field. 
A measurement of the joint state of the field and the unknown state $\ket{\psi}$ of qubit $A$, now leads to a random outcome and an accompanying measurement back action on qubit $B$. Neither the measurement outcome nor the projected state of qubit $B$ reveal any property of $\ket{\psi}$. But, the measurement outcome can be communicated classically to select and implement a local unitary on qubit $B$ which finally prepares it in the state $\ket{\psi}$.

 \begin{figure}
\includegraphics[width=\linewidth]{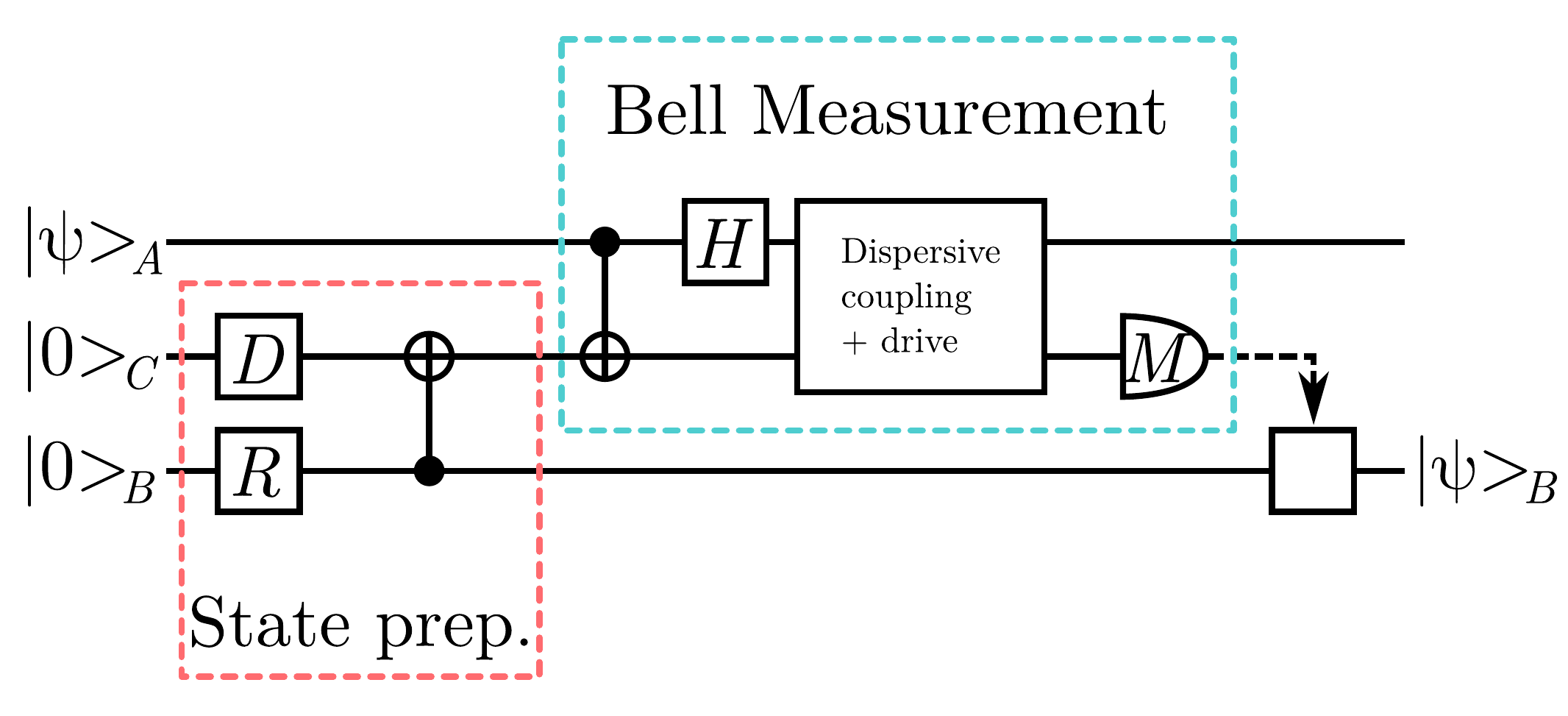}
\caption{Quantum circuit implementing teleportation of an unknown state of a qubit $A$ to another physical qubit $B$. System $C$ is a harmonic oscillator, initially entangled with qubit $B$: a $B$ qubit superposition state, prepared by the $\pi/2$-rotation $R$, interacts dispersively with a coherent state $|\beta\rangle_C$, prepared by the displacement operator $D$. The symbol in the figure indicates that resulting, conditional change of phase of the coherent state amplitude to $\ket{-\beta}_C$ is equivalent to C-NOT gate on the effective oscillator qubit states. A Bell measurement in the discrete basis of $A$ qubit states and coherent states $|\pm \beta\rangle_C$  is accomplished by a dispersive interaction between the systems followed by probing the field leaking from the cavity and then dispersively probing the state of the qubit $A$ in a rotated (by $H$) basis. Finally qubit $B$ is subject to a unitary operation depending on the outcome of the Bell measurement. The protocol is explained in detail in the text.} \label{fig:circuit}
\end{figure}

So far, the description of our proposal follows the protocols applied in teleportation experiments between two ions, coupled sequentially to a third ion \cite{riebe2004deterministic,barrett2004deterministic} and between superconducting qubits, coupled sequentially to a third superconducting qubit \cite{steffen2013deterministic},
with the main differences being that the communication channel is a different physical system, extending over both locations $A$ and $B$ (but decoupled by frequency detuning of the components).

While a cavity field offers continuous variable teleportation \cite{RevModPhys.74.145}, to teleport the discrete states of qubits we employ the so-called hybrid quantum teleportation schemes \cite{takeda2013deterministic} where a qubit degree of freedom is associated with a pair of coherent states of the field. For the initial entanglement generation we use that a moderately detuned qubit causes a qubit-state dependent frequency shift of the cavity, equivalent to a frequency shift arising from a Kerr effect~\cite{PhysRevLett.115.180501,PhysRevLett.111.053601}.
While a Bell state measurement conventionally involves projective measurements on both subsystems, our procedure consist of sequential steps where we only probe the signal leaking from the cavity by homodyne detection \cite{PhysRevLett.95.060501}. The restriction of our readout mechanism to continuous homodyne probing rather than instantaneous projective measurements requires a signal analysis to infer the optimal local operation on the isolated target qubit $B$. In this article we compare the direct application of the accumulated signal as an approximate measurement of the state of the system to a Bayesian analysis which attempts to retrodict from the whole signal record what would have been the most likely outcome of a projective measurement at the beginning of the probing sequence. Such inference problems, accounting for our knowledge about the state of a quantum system at a (past) time $t$ are of the type addressed by the past quantum state (PQS) formalism \cite{PhysRevLett.111.160401}.

We note that our use of retrodiction in connection with teleportation is related to delayed-choice experiments and to implementation of entanglement swapping protocols, where the heralding of a certain entangled state was only done after the state had already been detected and consumed \cite{peres2000delayed, RevModPhys.88.015005}. In this, article however, we do not aim to herald past states, but we shall rather use the past quantum state formalism, to infer which local operation may best accomplish the teleportation protocol. We demonstrate how our protocol is implemented in circuit QED, where cavities and qubits are constructed by superconducting waveguides and Josephson junctions respectively~\cite{wallraff2004strong, PhysRevA.69.062320}, but the analysis is general and can in principle be implemented in any system where the readout out of one or some of the components occurs sequentially and simultaneously with unitary or dissipative dynamics. The improvement over a more straightforward measurement and feed-forward protocol is the main result of this work.

The paper is organized as follows: In Sec. \ref{sec2} we describe our physical model and teleportation scheme. In Sec. \ref{sec3} we simulate the dynamics during continuous probing of the system and we present a simple signal analysis for the teleportation protocol. In Sec. \ref{sec:pqs} we apply the past quantum state analysis to (simulated) measurement data, and we show that we can obtain better teleportation fidelities from this approach. We present our conclusions and an outlook in Sec. \ref{sec:conc}.

\section{The protocol}
\label{sec2}

The protocol that we are proposing is depicted in Fig.~\ref{fig:circuit}. We consider two qubits and one cavity. At the beginning, one of the qubits, $A$, is in the unknown state, $\ket{\psi}_A$, that we intend to teleport. The second qubit, $B$, and the cavity are in the qubit ground state $\ket{0}_B$ and the oscillator ground state $\ket{0}_C$. We use the subscripts $A$, $B$, $C$ for qubit $A$, qubit $B$ and cavity respectively.

The teleportation protocol consists of three steps \cite{PhysRevLett.70.1895}. First we create an entangled state between the qubit $B$ and the cavity (a qubit-cavity Bell state) \cite{brune1992manipulation, vlastakis2013deterministically, vlastakis2015characterizing}. Then, we perform a Bell state measurement of $A$ and $C$, involving an entangling gate operation and probing of the field amplitude, followed by a field measurement that is sensitive to the qubit state.  The third and final step is the application of a unitary operation on the qubit $B$  chosen according to the outcome of the field measurements. Ideally, the combined protocol returns qubit $B$ in the desired state $\ket{\psi}_B$.

We encode a qubit degree of freedom in a pair of coherent states $\ket{\pm\beta}_C$ of the cavity oscillator, and we prepare a Bell state between $B$ and $C$ by first exciting the oscillator coherent state $\ket{\beta}_C$ and a qubit $B$ superposition state: $(\sqrt{2})^{-1}(\ket{0}_B+\ket{1}_B)$, followed by a dispersive interaction between the two systems which implements a qubit-controlled phase shift of the coherent state amplitude,
\begin{align*}
\ket{\phi}_{BC} &= U_{BC}\frac{1}{\sqrt{2}} (\ket{0}_B \ + \ket{1}_B )\ket{\beta}_C \\ &= \frac{1}{\sqrt{2}} (\ket{0}_B \otimes \ket{\beta}_C + \ket{1}_B \otimes \ket{-\beta}_C ).
\end{align*}
The operation $U_{BC}$ is equivalent to a C-NOT gate on the effective qubit subspace of the oscillator spanned by $\ket{\pm\beta}_C$ and the resulting state is equivalent to a two-qubit Bell state in the limit of a vanishing overlap of the two coherent states, $\braket{-\beta}{\beta} = \text{exp}(-2|\beta|^2) \approx 0$. We shall apply $\beta \simeq 2$ in our simulations resulting in a non-zero overlap and a correspondingly reduced fidelity of our protocol. We emphasize that after the state, $\ket{\phi}_{BC}$, is generated, qubit $B$ is not interacting with the cavity any further and the protocol is formally equivalent to true teleportation among spatially separated quantum systems.

The full system is now in a product state of the unknown qubit state  $\ket{\psi}_A= \alpha_0\ket{0}_A+\alpha_1\ket{1}_A$, and the entangled Bell-like state of the qubit $B$ and the cavity $C$, $\ket{\phi}_{BC}$, and following the teleportation procedure a Bell state measurement is performed on qubit $A$ and the cavity $C$. The Bell measurements can be decomposed into the application of a C-NOT gate and a qubit $A$ rotation, followed by qubit projective measurements on the separate systems $A$ and $C$. By expanding the state:
\begin{align}
H_AU_{AC}\ket{\psi}_A\ket{\phi}_{BC}&=\frac{1}{2}[\ket{0}_A(\alpha_0\ket{0}_B+ \alpha_1\ket{1}_B)\ket{\beta}_C\nonumber\\ &+\ket{0}_A(\alpha_0\ket{1}_B+\alpha_1\ket{0}_B)\ket{-\beta}_C\nonumber\\ &+
\ket{1}_A(\alpha_0\ket{0}_B-\alpha_1\ket{1}_B)\ket{\beta}_C\nonumber\\ &+\ket{1}_A(\alpha_0\ket{1}_B-\alpha_1\ket{0}_B)\ket{-\beta}_C].
\end{align}

We see, that depending on the measurement on qubit $A$ and the cavity, the conditional state of qubit $B$ acquires the unknown state amplitudes of the input state, and      we can immediately read out the conditional mapping between the qubit $B$ output state and the input state:
\begin{align}
0_A,\beta_C\mapsto\alpha_0\ket{0}_B+\alpha_1\ket{1}_B&=\ket{\psi}_B \nonumber \\
0_A,-\beta_C\mapsto\alpha_0\ket{1}_B+\alpha_1\ket{0}_B&=\sigma^x_B\ket{\psi}_B \nonumber \\
1_A,\beta_C\mapsto\alpha_0\ket{0}_B-\alpha_1\ket{1}_B&=\sigma^z_B\ket{\psi}_B \nonumber \\
1_A,-\beta_C\mapsto\alpha_0\ket{1}_B-\alpha_1\ket{0}_B&=-i\sigma^y_B\ket{\psi}_B \label{eq:teleportTab},
\end{align}
expressed in terms of the single-qubit Pauli operators $\sigma^x,\sigma^y,\sigma^z$.

\begin{figure}[b]
\centering
\includegraphics[scale=0.4]{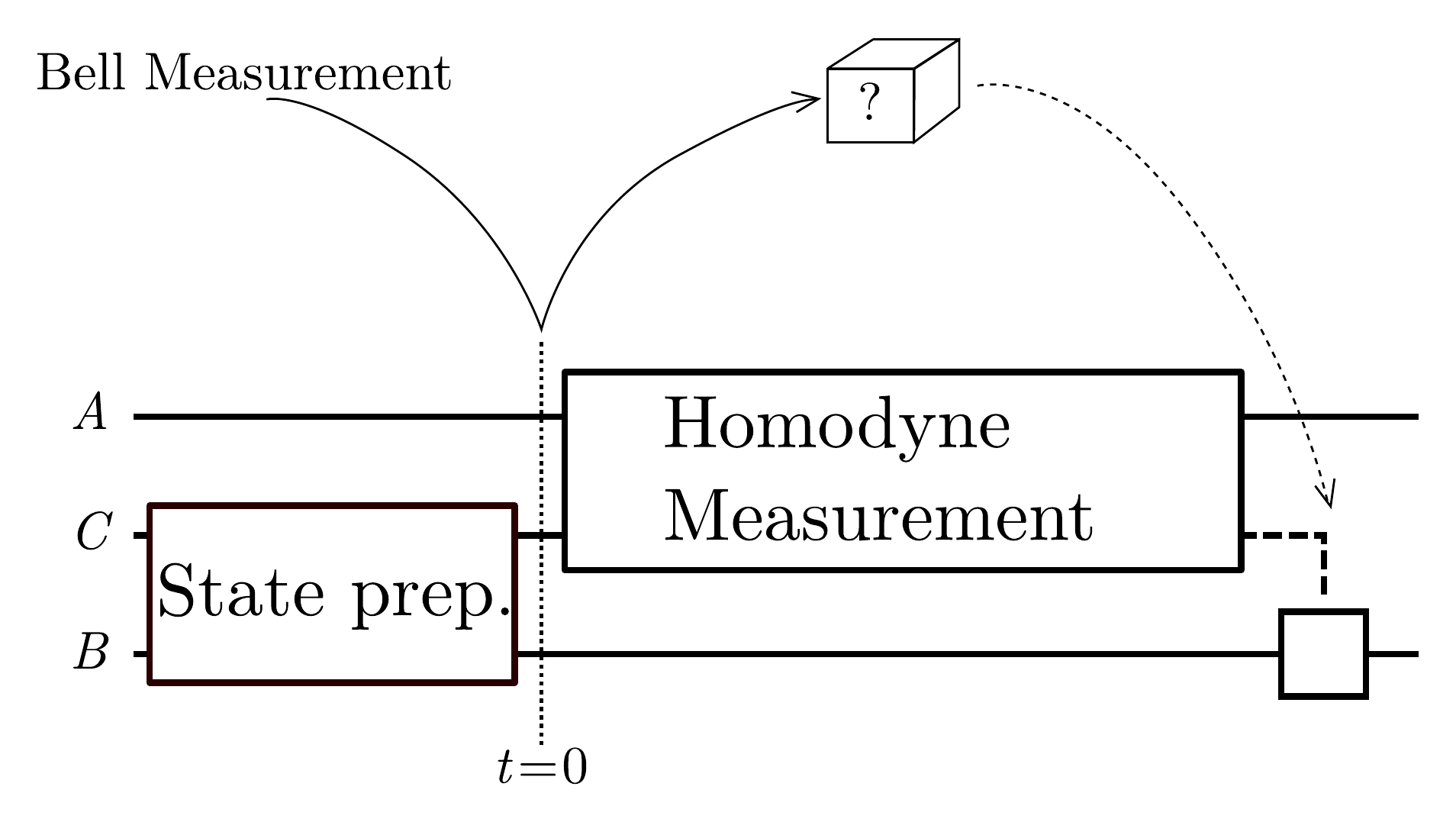}
\caption{Schematic representation of the time dependent measurement procedure. Our implementation uses continuous homodyne observation of the field emitted by the cavity, and after preparation of the entangled and input states and the controlled unitary operations on the input qubit and the cavity, we read out first the leaking cavity field amplitude and, subsequently, we drive the cavity and use the dispersive interaction to probe the state of qubit $A$.  Both measurements are subject to noise and they take a finite duration, and in the text we compare two analyses of the signals leading to different choices for the optimal operation on the undisturbed target qubit $B$: a direct approach based merely on the integrated signals, and an approach using the past quantum state formalism to retrodict what would have been the most likely projective Bell measurement outcome at the earlier time $t=0$.}
\label{fig:bell}
\end{figure}

The four Bell state measurement outcomes \eqref{eq:teleportTab} occur with equal probabilities, independently of the input state, and they hence reveal no information about the teleported state. In the experiment, one merely has to perform the measurements, and the combination of the measurement back action on qubit $B$ due to the entanglement with the cavity and the conditional unitary operation, cf., \eqref{eq:teleportTab} should ensure the correct state transfer.

Now, in the architecture, presented in the next section, we restrict ourselves to continuous homodyne detection of the output signal from the resonator. Such measurement can be used to first distinguish the coherent states, $\ket{\pm\beta}$, as the field leaks from the cavity and, subsequently, we can drive the cavity with a resonant input field, which undergoes a phase shift due to the interaction with qubit $A$ and which hence permits detection of the qubit state~\cite{PhysRevLett.95.060501}. Both measurements are noisy and take time, but since qubit $B$ is not addressed during the probing, we assume that the optimal unitary operation on $B$ after the measurement is complete, is the one pertaining to our best estimate of which Bell state was occupied at the initial time of the measurement ($t=0$ in Fig.~\ref{fig:bell}). To assess the fidelity of our protocol, we choose random input
states, simulate the protocol and, in particular, the probing, and we check how well our different strategies succeed on average in teleporting the state. Note, however, that since our strategies should not invoke any knowledge of the input state, we are not allowed to use any properties of the conditioned quantum state in the choice of operations on the system. We shall thus design the procedure to depend only on the simulated signal (equivalent to an actual measured signal in an experiment).

There has been a number of successful proposals for tests and analyses of the teleportation protocol based on the study of the density matrix describing the system as a whole, i.e. including the entangled pair and the state to be teleported \cite{PhysRevA.66.022316,PhysRevLett.91.257903,1464-4266-5-3-370, PhysRevA.84.052327}. The evaluation of a forward evolution of a quantum master equation, however, requires prior knowledge of the initial state, which is not known. This can be compensated by averaging over all possible initial state or by adding an ancillary degree of freedom \cite{PhysRevA.54.2614}. Our approach, on the other hand, does not presuppose anything about the state to be teleported and is experimentally reproducible for any state $\ket{\psi}_A$.

\section{Implementation and simulation}
\label{sec3}

\begin{figure}[b]
\centering
\includegraphics[scale=0.3]{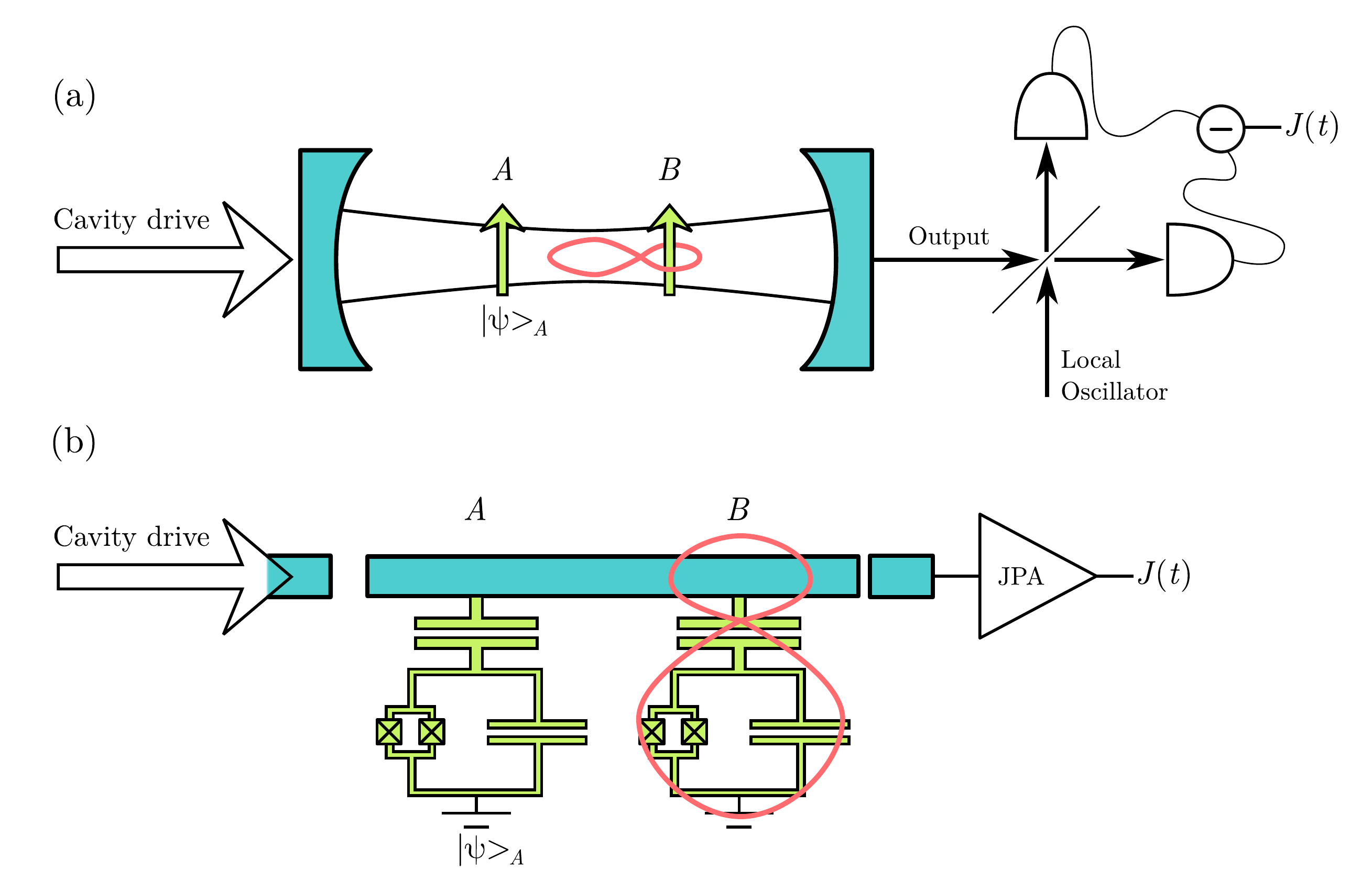}
\caption{Schematic presentation of the experimental setup. Panel (a) depicts a quantum optics rendering of the experiment with optical homodyne detection, while panel (b) shows the scheme for superconducting circuits with two transmon qubits, one resonator cavity and a Josephson parametric amplifier (JPA) for amplification and measuerement of a field qudrature variable.}
\label{fig:scheme}
\end{figure}

We shall implement the protocol with two qubits and a cavity  field, illustrated in a quantum optics and in a circuit QED schematic in Fig. \ref{fig:scheme}. A two-level system coupled to a cavity field is described by the Jaynes-Cummings Hamiltonian,
\begin{align}
H = \omega_r a\dag a + \frac{\Omega_q}{2} \sigma_z + g (a\dag \sigma_- + a \sigma_+),
\end{align}
where $a (a^{\dagger})$ is the annihilation (creation) operator for the field in the cavity, $\sigma_{-(+)}$ is the lowering (raising) operator for the two-level system, $\omega_r$ is cavity frequency, $\Omega_q$ is the qubit frequency and $g$ is the coupling strength.
This Hamiltonian has been demonstrated with numerous effective two-level systems coupling to cavity fields, e.g. atoms~\cite{haroche2006exploring}, quantum dots~\cite{reithmaier2004strong} and in particular superconducting qubits in circuit QED~\cite{wallraff2004strong, PhysRevA.69.062320}. In the regime where the detuning between the cavity and qubit frequencies, $\Delta = \Omega_q - \omega_r$, is much larger than the coupling strength, $g$, we arrive at the dispersive Hamiltonian
\begin{align}
H = \omega_r a\dag a + \frac{\Omega_q}{2} \sigma_z + \chi \, \sigma_z a\dag a,
\end{align}
with the dispersive coupling $\chi = g^2/\Delta$. In circuit QED, the two-level systems are superconducting qubits and the cavity field is the microwave field of a coplanar waveguide or the field within a 3D microwave cavity. Circuit QED also makes it possible to tune the frequency of qubits and thereby the dispersive coupling $\chi$ and similarly a transmon with tunable coupling, $g$, can be constructed~\cite{PhysRevLett.106.030502, PhysRevLett.106.083601}.  In the relevant regime of transmon qubits \cite{PhysRevA.76.042319}, $\chi$, attains a form similar to the result for general two-level systems (see Appendix \ref{app:transmon}).

Considering the scheme presented in Sec. \ref{sec2} we have two qubits dispersively coupled to the cavity field. We initialize the system such that \mbox{$\chi_A = \chi_B \approx 0$} (see Appendix \ref{app:transmon}). A coherent displacement of the cavity field is achieved by applying a coherent drive, \mbox{$H_d = \epsilon_d(t) a\dag + \epsilon_d^*(t) a$}, to the cavity. Similarly, a single qubit rotation can prepare the qubit in an arbitrary
superposition state, as required for the input state and
for the protocol. Tuning qubit $B$ close to resonance with the cavity yields a non-zero dispersive coupling, $\chi_B > 0$. This results in the application of a phase gate, $U_{BC}$, when the dispersive interaction is active for a time $\tau = \pi/\chi_B$ \cite{vlastakis2015characterizing} (details presented in Appendix \ref{app:transmon}), and we obtain the Bell-state required for the teleportation protocol. The same gate is equivalently applied for qubit $A$ later in the protocol, where qubit $B$ is far-detuned from the cavity and, thus, well-separated, albeit in frequency space rather than in physical space.

For the readout we leave both qubits far-detuned from the cavity, and we monitor the signal leaking from the cavity using a quantum-limited parametric amplifier \cite{PhysRevA.39.2519, castellanos2007widely, PhysRevB.83.134501, lin2013single} which performs a highly efficient homodyne detection. As we are not directly measuring the intra cavity field quadrature $X_c = a + a\dag$ but the radiation emitted over time, we describe the dynamical evolution of the homodyne measurement by the stochastic master equation~\cite{WiMi2010}
\begin{align}
d\rho &= -i [H ,\rho]\,dt + \kappa \Big(a \rho a\dag - \frac{1}{2}(a\dag a \rho + \rho a\dag a) \Big)dt \nonumber \\
&\quad + \sqrt{\kappa \eta} \Big( a \rho + \rho a\dag - \text{Tr}(X_c\rho) \, \rho \Big)\,dW(t), \label{eq:homomaster}
\end{align}
with $\kappa$ being the linewidth of the cavity and $dW(t)$ being a stochastic Gaussian process with $\exv{dW^2} = dt$ and $\exv{dW} = 0$ which represents the stochastic
part of the homodyne measurement signal
\begin{align}
J(t) = \sqrt{\kappa \eta} \, \text{Tr}(X_c\rho) + \frac{dW(t)}{dt},
\end{align}
which directly corresponds to the experimentally available current. The stochastic part is zero in the mean and averaging $J(t)$ over time yields the quantity
\begin{align}
\label{SbetaFWD}
S_{\beta} = \text{sign}\Big(\int_0^{T_\beta} J(t)\, dt \Big).
\end{align}
which can be used to distinguish the two states $\ket{\beta}$ and $\ket{-\beta}$ as they lead to mean amplitude signals with opposite sign. The integral \eqref{SbetaFWD} includes a noise component which is dominant for short integration times, and the duration $T_\beta$, devoted to distinguish $\pm\beta$ should be chosen on the order of the cavity lifetime $\kappa^{-1}$, after which the signal has decayed.

As a next step, the frequency of qubit $A$ is tuned close to the cavity resonance and a non-zero dispersive shift is obtained, that we can probe via the phase of a transmitted coherent drive with constant amplitude $\epsilon_r$. The mean-field coherent state amplitude of the cavity field now follows the equation
\begin{equation}
\frac{d}{dt}\beta=-i\chi_A\langle\sigma_z\rangle\beta - \frac{\kappa}{2}\beta - i\epsilon_r,
\end{equation}
where $\chi_A$ is the dispersive shift. The steady state solution ($d\beta_{ss}/dt=0$) yields
\begin{align}
\beta_{ss} = \frac{ -i \,\epsilon_r\kappa- 2\epsilon_r \chi_A \exv{\sigma_z}}{\frac{\kappa^2}{2} + 2\chi_A^2 \exv{\sigma_z}^2}
\end{align}
with the real part, probed by the JPA,
\begin{equation}
\textup{Re}(\beta_{ss})=\frac{ - 2\epsilon_r \chi_A \exv{\sigma_z}}{\frac{\kappa^2}{2} + 2\chi_A^2 \exv{\sigma_z}^2}.
\end{equation}
The sign of the average integrated signal is in steady state governed by the sign of $\langle\sigma_z\rangle$, i.e., the qubit state. To avoid transient contributions without a definite sign to the signal, we accumulate the probe signal after a finite waiting time $T_w$, that we shall identify by numerical optimization,
\begin{align}
\label{SaFWD}
S_{A} = \text{sign}\Big(\int_{T_\beta+T_w}^{T_\beta + T_m} \hspace{-18pt} J(t)\, dt \Big).
\end{align}
A negative sign of this integrated current implies the excited state qubit $\ket{1}$, while a positive sign corresponds to the ground state $\ket{0}$. The measurement may be in error due to the Gaussian noise contribution to the integrated signal, but  the readout is a quantum non-demolition measurement, and increasing $T_m$ will increase the measurement fidelity until we approach the qubit lifetime.

Knowing the time dependent mean field envelopes during the two probing periods allows accumulation of the signal with a time dependent weight factor. We have implemented such weighing, but for our parameters we did not see a significant change of the resulting teleportation fidelity, beyond what we achieve by optimizing the parameters $T_\beta, T_m$ and $T_w$.

\section{Past quantum state simulation}
\label{sec:pqs}

The past quantum state (PQS) formalism is a generalization of classical smoothing algorithms and of the so-called forward-backward analysis of hidden Markov models \cite{baum1966, PhysRevA.89.043839} to the case of quantum states. The aim of the formalism is to infer from measurements on quantum system before and after a given time $t$ what would have been the outcome of an arbitrary measurement if carried out at time $t$. It may also be viewed as a generalization of the Aharonov, Bergmann and Leibowitz (ABL) rule, for projective measurements performed between the preparation and post-selection of a quantum system in arbitrary pure states \cite{ABL}, and of the theory of weak value measurements for the post selected average of weak, non-projective measurements \cite{PhysRevLett.60.1351}. It exactly reproduces the results of these theories, but for our case of dynamically evolving and continuously monitored quantum systems we shall need the more general theory of \cite{PhysRevLett.111.160401}.

Conventionally, our knowledge about a quantum system subject to damping, Hamitonian evolution and measurements is governed by a density matrix $\rho(t)$, which may be calculated by use of a stochatic master equation that incorporates the effect of measurements prior to $t$, cf., Eq.~(\ref{eq:homomaster}).
This density matrix, $\rho(t)$, yields the probability for the outcome of any measurement on the system as represented most generally by positive operator valued measures \cite{nielsen2010quantum} (POVMs), $\{M_i\}$, with $\sum_iM_i^{\dagger}M_i=\mathbbm{1}$, for which outcome $i$ occurs with the probability, $p_\rho(i)=\textrm{Tr}(M_i\rho M_i^\dagger)$.

The further knowledge due to later measurements is represented by another matrix $E(t)$, which together with $\rho(t)$ yields the POVM outcome probabilities, conditioned on all, earlier and later measurements on the system \cite{PhysRevLett.111.160401, xu2015correlation},
\begin{equation}
\label{eq:Born}
p(i)=\frac{\tr{M_i\rho M_i^{\dagger}E}}{\sum_{i}\tr{ M_i\rho M_i^{\dagger}E}}.
\end{equation}
This expression follows from a quantum mechanical analysis of the measurement situation \cite{PhysRevLett.111.160401} and it has important qualitative and quantitative consequences as illustrated and verified in recent experiments with atoms and with superconducting qubits \cite{PhysRevA.91.062116, PhysRevLett.114.090403,PhysRevLett.112.180402}.

In practical calculations we find $E(t)$ by solving a stochastic master equation backwards in time from the last moment, $T$, of measurements, where $E(T)=\mathbbm{1}$.
The master equation for $E$ is the adjoint to the master equation for $\rho(t)$, and it involves the same measurement current signal $J(t)$ as applied for the calculation of $\rho(t)$. While $\rho(t)$ only depends on $J(t')$ for $t' \leq t$, $E(t)$ depends on $J(t')$ for $t' \geq t$.
From Eq.~(\ref{eq:homomaster}), we thus obtain
\begin{align}
\label{eq:MasterBckw}
d{E}  &= i[H,E]\,dt + \kappa \Big(a\dag E a - \frac{1}{2}(a\dag a E + E a\dag a) \Big)dt \nonumber \\ &\quad + \sqrt{\eta \kappa} J(t) \Big( a\dag E + E a \Big)\,dt.
\end{align}
Note that a $c$-number term is missing compared to Eq.(\ref{eq:homomaster}). This will only affect a common factor on $E$, but since the adjoint master equation is not trace preserving anyway, and since Eq. \eqref{eq:Born} explicitly renormalizes the probabilities, this does not affect retrodictions made by the theory.

Our aim is to infer which action to apply on the target qubit. Under ideal circumstances this action is inferred from a projective Bell state measurement at time $t=0$, but we are not able to perform that measurement. As an alternative candidate procedure, we shall employ the past quantum state formalism to infer from our continuous measurements until the later time $T = T_\beta+T_m$ what would have been the most likely outcome of a projective Bell measurement, if it had taken place at $t=0$.

Without knowing the input state $\ket{\psi}_A$, our prior knowledge about the outcome is equivalent to a fully mixed density matrix $\rho(t=0)\propto \mathbbm{1}/4$ on the tensor product state space of $\ket{0}_A,\ \ket{1}_A$ and $\ket{\beta}_C,\ \ket{-\beta}_C$.  We calculate $E(0)$  by solving the  backward evolution Eq.~\eqref{eq:MasterBckw} from $T=T_\beta + T_m$ to 0  (note that we need only to solve this equation for the cavity and qubit $A$, as qubit $B$ is a passive spectator under the homodyne detection sequences). Finally, to retrodict the most likely Bell state outcome, the gates are performed according to Fig.~\ref{fig:circuit} and then we need the outcome of a qubit $A$ projective measurement, and a measurement of the sign of the intra cavity field quadrature (see Eq.~\eqref{eq:teleportTab}).

This measurement is described by the POVM
\begin{equation}
\label{BellPovm}
M_i=\ket{i}_A\ket{X}_{C C}\bra{X}_A\bra{i},
\end{equation}
where $X=\pm\beta$ and $i=0,1$. This, together with the backward evolution \eqref{eq:MasterBckw}, gives us the tools for evaluating the generalized Born rule \eqref{eq:Born} at the time zero and therefore calculate the outcome probabilities for the 4 different outcomes of the Bell measurement \eqref{BellPovm}.

\begin{figure}
\centering
\includegraphics[scale=1.0]{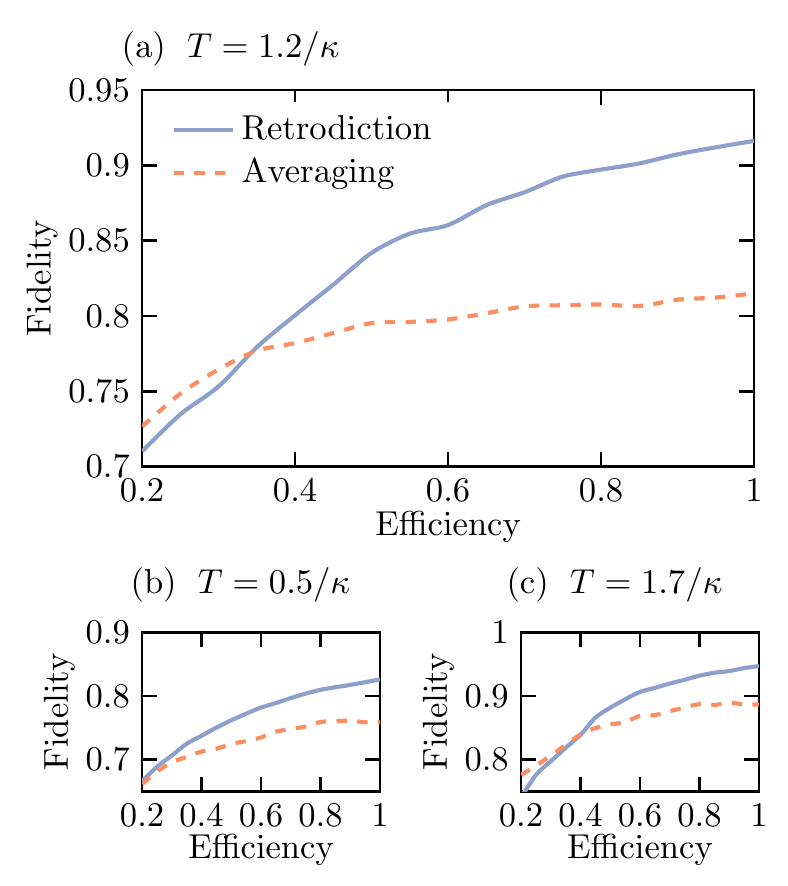}
\caption{The fidelity of the protocol as a function of efficiency pre- and post the past quantum state retrodiction is applied. Here the dashed red curve is the results obtained used the avereging in Eqs.~\eqref{SbetaFWD}~and~\eqref{SaFWD}, while the solid blue curve is the fidelity obtained using the past quantum state analysis to estimate the most like Bell-measurement outcome. The parameters used in the simulations are explained in the text.}\label{fig:fidelity}
\end{figure}

Having used the PQS formalism to retrodict what would have been the most likely Bell state outcome, in every run of the experiment (simulation), we choose to apply the corresponding unitary corrective operation on qubit $B$ as in Eq.~\eqref{eq:teleportTab}. That state, in turn, can be directly compared with the randomly chosen input state, and we can determine the average fidelity of the protocol as function of the physical parameters.

\section{The results}
\label{sec:results}

We carry out numerical simulations for a cavity with a damping rate $\kappa$, initially prepared in a coherent state with $\beta = 2$. We assume a dispersive interaction with $\chi = 13.5 \kappa$, and a coherent drive of $\epsilon_r = 2\chi$ during the qubit readout. We measure for a time $T = T_\beta + T_m$ and we fix $T_\beta / T = 0.4$. To represent real experiments, we also perform simulations with different values of the detector efficiency.  With these parameters, the signal-to-noise ratio is expected to only exceed unity for $T$ larger than $1/\kappa$ \cite{RevModPhys.82.1155}. The forward evolution of the conditioned  master equation, Eq.~\eqref{eq:homomaster}, yields the state of the entire system, and after application of the unitary operation $\sigma_B$, chosen according to Eq.~\eqref{eq:teleportTab},  we obtain the reduced density matrix of the target qubit, $\rho_{B} = \text{Tr}_{AC}[\sigma_{B}^i\rho(T)\sigma_{B}^i]$.
The overlap, $_A\bra{\psi}\rho_{B}\ket{\psi}_A$, yields the fidelity of the final state and we estimate the protocol fidelity as the average over the state fidelities \cite{PhysRevA.66.022316,PhysRevLett.91.257903} by choosing 500 random input states from a uniform distribution over the Haar measure~\cite{JW14Haar}.
\begin{figure}
\centering
\includegraphics[scale=1.0]{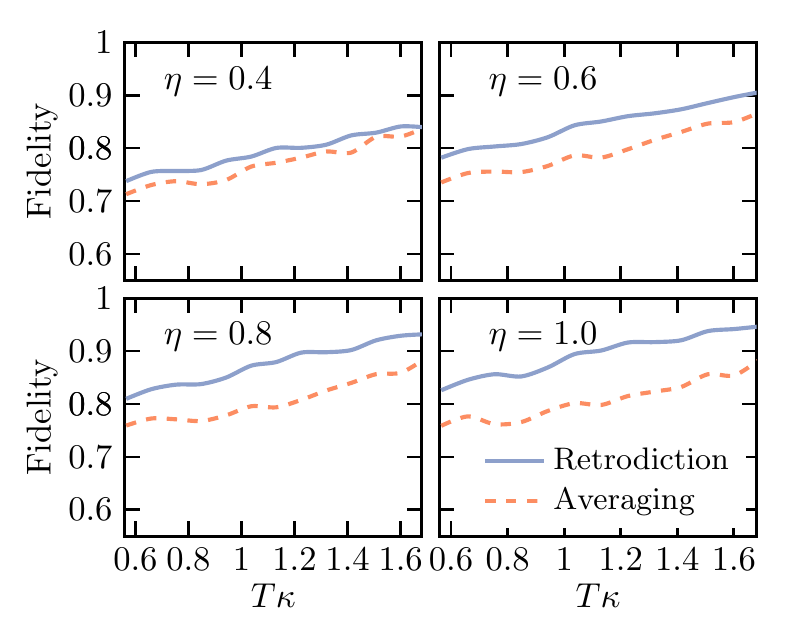}
\caption{The fidelity of the protocol as a function of the total measuring time after the gates were applied. Each panel displays the fidelity of the protocol for the efficiency shown in the panel with the parameters explained in the main text.}\label{fig:time}
\end{figure}
The results of such simulations are shown as function of the detector efficiency in Fig. \ref{fig:fidelity} and as function of the probing time in Fig.~\ref{fig:time}, where the red dashed lines show the fidelity of the procedure when $\sigma_B$ is chosen from the values of the integrated homodyne signals and the solid blue lines show the result when the past quantum state retrodiction is applied. For low detector efficiency and for short probing times, the methods have comparable and rather low fidelities, while for higher efficiencies and longer probing times, the past quantum state protocol generally performs better.  In Fig. \ref{fig:fidelity} (b) we see that the short readout time does provide enough data for more than a marginal improvement while the long readout time in Fig. \ref{fig:fidelity} (c), accumulates enough data that also the integrated signal allows a good estimation of the optimal $\sigma_B$, and hence there is less room for improvement. Our interest in the intermediate probing time in Fig.~\ref{fig:fidelity} (a), where the past quantum state offers the largest improvement, is motivated by the existence of incoherent processes in the qubit $B$, which is not included in the present analysis but which will become important for long probing times.

\section{Outlook and Conclusion}
\label{sec:conc}
We have presented a scheme for quantum teleportation between two qubits taking turns interacting with a resonant cavity mode. Subsequent application of a continuous, homodyne measurement of the signal leaking from the cavity constitutes an approximate Bell state measurement. The accomplishment of this readout is significantly restricted by the measurement noise. Using a Bayesian analysis, in the form of the past quantum state formalism, we show that it is possible to increase the fidelity of the protocol over schemes based on the value of the integrated signal alone. The protocol is not exact, and its high fidelity limit merges naturally with the limit where also the integrated signal yields a successful outcome. We thus imagine that the main improvement will be for parameters where, e.g., decoherence and decay of the qubit components forbid high signal-to-noise probing of the system. Our results suggest that similar detailed analyses may improve other protocols implemented in leaky continuously monitored quantum systems.

Our protocol was exemplified in a circuit QED setup with an experimentally motivated preparation of the initial entangled state. We note, however, that the preparation of this initial state is not unique. For example, state-of-the-art remote entanglement schemes \cite{PhysRevLett.112.170501, 2016arXiv160303742N} could be adapted to perform the state preparation. Similarly, after the state preparation, the target qubit state may be transferred to travelling microwave photons \cite{PhysRevB.84.014510, PhysRevLett.112.210501, PhysRevX.4.041010} without changing the rest of the protocol presented here. Moreover, the method can be easily translated to any other platform with a cavity and two-level systems. Our technique can be generalized to experimental realization of entanglement swapping \cite{2007EntanglementSwapping}, where a mixed state (one half of maximally entangled pair in particular) is teleported instead of a pure one.

Admittedly, quantum teleportation as a quantum protocol was originally formulated with the state transfer over very long distances in mind. In a circuit QED setting we can only teleport the state between two qubits with the relatively small state separation inside of the resonator. However, the length of a resonator can be up to centimeters, so it can still cover considerable distance on a chip. A network of resonators and qubits could be used for implementation of a quantum repeater \cite{RevModPhys.83.33} for arbitrary entanglement swapping on the large chip. While there are other techniques available for remote entanglement distribution \cite{PhysRevA.92.032308,PhysRevLett.112.170501,2016arXiv160303742N},
the teleportation scheme presented here provides a tool to explore a much wider range of quantum computation and communication protocols in a circuit QED system. Quantum teleporation for instance provides a universal resource for quantum computing~\cite{gottesman1999demonstrating} and plays a role in quantum error correction schemes~\cite{lidar2013quantum}, in particular in measurement based quantum computing~\cite{PhysRevLett.86.5188, leung2001two, nielsen2003quantum}. Teleportation based techniques are also essential for modular approaches to both distributed quantum computing~\cite{devoret2013superconducting} and quantum error correction~\cite{li2015hierarchical}.

\section*{Acknowledgements}
The authors acknowledge financial support from the Villum Foundation Centre of Excellence, QUSCOPE. Furthermore, CKA acknowledges support from the Danish Ministry of Higher Education and Science.

\begin{appendix}

\section{Tunability and gates with transmon qubits}
\label{app:transmon}
A transmon qubit is a superconducting qubit consisting of a Josephson junction shunted with a large capacitance. The Hamiltonian describing the transmon is given as \cite{PhysRevA.76.042319}
\begin{align}
H = 4E_C \hat{n}^2 - E_J \cos \hat{\phi} \label{eq:transmon}
\end{align}
with $\hat{n}$ the Cooper-pair number on the superconducting island, $\hat{\phi}$ the phase drop across the junction and the Josephson energy, $E_J$ set much larger than the charging energy, $E_C$. Due to the anharmonicity of the cosine potential we can restrict the dynamics to the two lowest lying states, such that we have the qubit Hamiltonian $H = \Omega \sigma_z /2$ with $\Omega = \sqrt{8E_J E_C}-E_C$ ($\hbar=1$). The transmons are typically capacitively coupled to a transmission line resonator such that we have the Jaynes Cummings Hamiltonian
\begin{align}
H = \omega_r a\dag a + \frac{\Omega}{2} \sigma_z + g(a\dag \sigma_- + a\sigma_+)
\end{align}
which in the off-resonant regime, $|\Delta| \gg g$ with $\Delta = \Omega - \omega_r$, translates into the dispersive Hamiltonian
\begin{align}
H \approx \omega_r a\dag a + \frac{\Omega}{2} \sigma_z + \chi \sigma_z \, a\dag a.
\end{align}
The transmon is, however, not a true two-level system; the Hamiltonian \eqref{eq:transmon} supports many levels and the energy difference between the second transition frequency, $\Omega_{21}$, and the first can be found to be
\begin{align}
\alpha = \Omega_{21} - \Omega = -E_C.
\end{align}
The third level therefore influence the off-resonant coupling and we can calculate the dispersive shift \cite{PhysRevA.76.042319}
\begin{align}
\chi = - g^2 \frac{E_C}{\Delta (\Delta - E_C)}.
\end{align}
Let us now show how to use this coupling for the controlled phase gate used in the main text. For this gate, we need $\chi$ to be tunable, which can be achieved by using a tunable $g$ or by tuning the frequency. The frequency can be tuned simply by replacing the single Josephson junction with a SQUID, such that the external flux through the SQUID changes the Josephson energy as $E_J(\Phi_x) = E_J(0) |\cos \Phi_x|$.

For our scheme we would have a resonator with a larger frequency than the qubits and we initially park the qubits at a flux $\Phi_x^0 > 0$, such that we have a very large detuning. By slowly changing the flux, we can tune the frequency e.g. as a Gaussian pulse,
\begin{align}
\Delta(t) = \begin{cases} \Delta_0 - \Delta_t e^{-(t-t_0)^2 / \tau^2} & \text{ for } t<t_0 \\ \Delta_0 - \Delta_t & \text{ for } t_0 \leq t \leq t_1 \\ \Delta_0 - \Delta_t e^{-(t-t_1)^2/\tau^2} & \text{ for } t_1 < t, \end{cases}
\end{align}
with $\Phi_x(t) = 0$ for $t_0 \leq t \leq t_1$. Now, the accumulated phase on a coherent state in the resonator will be
\begin{align}
\phi(t_1-t_0) &= \int_{t_0-5\tau}^{t_1+5\tau} 2\chi(t) \, dt \\&= - \int_{t_0-5\tau}^{t_1+5\tau}  2g^2 \frac{E_C}{\Delta(t) (\Delta(t) - E_C)} \, dt,
\end{align}
with the cut-off set at $5\tau$ and by choosing the waiting time $t_1 - t_0 = t_\pi$ we can find $\phi(t_\pi) = -\pi$, which will implement the phase gate.

To account for the parameters used in the simulation of the main text we can calculate all quantities using experimentally realistic values. We want to use a fairly high quality resonator so we use a linewidth of $\kappa = 2\pi \times 150$ kHz. For the qubit we fix the charging energy at $E_C = 2\pi \times 300$ MHz and use $E_J/E_C = 75$, which yields a qubit frequency of $2\pi \times 7.35$ GHz. We couple the qubit weakly to the resonator with a coupling $g = 2\pi \times 31$ MHz and fix the resonator frequency at $\omega_r - \Omega = 2\pi\times 250$ MHz. This yield a dispersive shift of $\chi = 2\pi\times 2.1$ MHz. We can now tune the flux to $\Phi_x^0 = 0.3\pi$, which will give $\Delta_t = 2\pi \times 1.9$ GHz and make $\chi$ much smaller than $\kappa$ when the qubit is set at the $\Phi_x^0$. For the phase gate we used a duration of $\tau = 10$ ns and we use the same duration when tuning the qubit into the readout position at the end of our scheme.

\end{appendix}

\bibliography{bt}

\end{document}